\DeclareRobustCommand{\baselinestretch{2}}

\documentclass[twocolumn,showpacs,preprintnumbers,amsmath,amssymb]{revtex4}

\usepackage{graphicx}
\usepackage{dcolumn}

\def\p1{\partial_1}
\def\p2{\partial_2}
\def\p3{\partial_3}

\begin{document}

\title{Time for pulse traversal through slabs of dispersive and negative ($\varepsilon$, $\mu$) materials}

\author{Lipsa Nanda, and S. Anantha Ramakrishna}

\affiliation{Department of Physics, Indian Institute of Technology Kanpur, Kanpur 208016, India}

\begin{abstract}

The traversal times for an electromagnetic pulse traversing a slab of dispersive and dissipative material with negative dielectric 
permittivity ($\varepsilon$) and magnetic permeability ($\mu$) have been calculated by using the average flow of electromagnetic energy in the medium. 
The effects of bandwidth of the pulse and dissipation in the medium have been investigated.
While both large bandwidth and large dissipation have similar effects in smoothening out the resonant features that appear due to Fabry-P\'{e}rot 
resonances, large dissipation can result in very small or even negative traversal times near the resonant frequencies.
We have also investigated the traversal times and Wigner delay times for obliquely incident pulses and evanescent pulses. The coupling to slab plasmon
polariton modes in frequency ranges with negative $\varepsilon$ or $\mu$ is shown to result in large traversal times at the resonant conditions.
We also find that the group velocity mainly contributes to the delay times for pulse propagating across a slab with $n=-1$. We have checked that
the traversal times are positive and subluminal for pulses with sufficiently large bandwidths.
\end{abstract}
\pacs{42.25.Bs, 42.65.-k, 78.20.Ci, 41.20.Jb}

%
\maketitle 
\section{Introduction}

The time for traverse of light through a dispersive medium is interesting and important, both from a fundamental
viewpoint\cite {chiao95,landauer94}, and for technological applications such as designing delay lines 
or systems for enhanced non-linear 
applications. However, there are a variety of time scales, depending on the physical quantity being measured, that
can be defined for this time of traverse\cite{chiao95,landauer94,hauge89,sar_epl}. 
A popular measure for the delay time of pulses is the Wigner delay time\cite{wigner}
$\tau_w = \frac {\partial \phi}{\partial \omega}|_{\omega=\bar {\omega}}$, 
i.e, the frequency derivative of the
phase of the output wave evaluated at the carrier frequency $\bar {\omega}$. The Wigner delay time,
which is based on tracking a feature on the pulse moving at the group velocity
($v_g=\frac {\partial \omega}{\partial k}$), can turn out to be superluminal or even negative, while at the same time
well describing the motion of pulses with narrow bandwidths over short distances\cite{garett,chiao94,wang00}. 
While there is no contradiction of Causality or of Special Relativity in these phenomena 
involving holomorphic pulses, the Wigner delay time
becomes inaccurate for large pulse bandwidths or when there is a large deformation in the pulses.

A measure that is based on the flow of electromagnetic energy for the time of traverse between two points
$\mathbf{r_i}$, and $\mathbf{r_f}$ was proposed by Peatross et al\cite{peatross}, which is given by,
\begin{equation}
\Delta t=\langle t\rangle_\mathbf{r_f}-\langle t\rangle_\mathbf{r_i},
\end{equation}
where
\begin{equation}
\langle t\rangle_\mathbf{r} = \frac{\mathbf{u}\cdot\int_{-\infty}^\infty t
\mathbf{S}(\mathbf{r},t) dt}{\mathbf{u}\cdot\int_{-\infty}^\infty
\mathbf{S}(\mathbf{r},t) dt}
\end{equation}
represents the arrival time of a pulse at a point $\mathbf{r}$.
This time scale is particularly suitable for pulses with large bandwidths as the relative contributions due to the
propagation at the group velocity (group delay), and deformation of the pulse (reshaping delay) can be
identified. Eq. (2) can be rewritten exactly as \cite{peatross} 
\begin{equation}
\langle t\rangle_\mathbf{r} =  \frac{\mathbf{u}
\cdot\int_{-\infty}^\infty \mathrm{Re}\left[-i\frac{\partial \mathbf{E}(\mathbf{r},\omega
)}
{\partial \omega} \times \mathbf{H}^{\ast}(\mathbf{r},\omega)\right] d \omega}
{\mathbf{u}\cdot\int_{-\infty}^\infty\mathbf{S}(\mathbf{r},\omega) d\omega},
\end{equation}
which is very useful for spectral calculations. This definition of the traversal time based on the motion
of the centroid of the Poynting vector has been supported by experiments on dispersive media\cite{tomita}, 
and even in angularly dispersive systems\cite{Aminul}. It has also been shown that the arrival times 
for pulses measured through the rate of absorption
in an ideal impedance matched detector is equivalent to the above arrival times\cite{lipsa}.
We have also shown earlier that the definitions of the group delay times, and the reshaping delay times 
get interchanged for evanescent pulses\cite{lipsa}.

Negative refractive index media (NRM) or left handed media (LHM) simultaneously have $\mathrm{Re}(\varepsilon)<0$, and 
$\mathrm{Re}(\mu)<0$ at a given 
frequency\cite{sar05}, and have captured the imagination of the Physics community by their numerous counter-intuitive
electromagnetic properties (See \cite{sar05} for a review of NRM).
In isotropic NRM, the wave vectors $\mathbf{k}$, and the Poynting vector $\mathbf{S}$ point in opposite directions.
NRMs are also necessarily dispersive and dissipative in nature, and can additionally also support surface (plasmons)
states on their interfaces with positive media. Thus the study of the phase velocity 
($v_p= \frac {\omega}{k}= \frac {c}{n}$), the group velocity, and the energy flow in such media is interesting, and
important. In fact in some metamaterials, one can have all possible combinations of (positive, and negative)
phase, and group velocities \cite{soukoulis_science06}.

In this paper, we study the times for pulse traversal through slabs of dissipative media with negative
material parameters ($\varepsilon < 0$, $\mu < 0$) using the average energy flow in the media given by \cite{peatross}.
Pulse propagation in NRM has been principally studied with a focus on negative refraction 
at interfaces \cite{pacheco_kong,lu_sridhar,foteinopoulou}, and 
nonlinear effects \cite{scaloraPRL05,aguannoPRL04,zharovPRL03}. 
The Wigner delay time has been studied for pulses at normal incidence on slabs of NRM in the limit of 
zero dissipation \cite{duttagupta}. We have earlier investigated the traversal times based on the energy flow in infinitely extended NRM, 
and for normally incident pulses in semi-infinite NRM \cite{lipsa}. 
There the geometry was chosen such that the traversal times were affected 
by only the intrinsic dispersion of the medium parameters, and avoided the effects of other resonances such as slab 
resonances (by using infinite or semi-infinite media), and surface plasmon resonances (by normal incidence of the pulses).
Here we will consider these effects of resonances on the traversal times for the transport of narrowband, and broadband
pulses through slabs of causal NRM. We also show that finite levels of dissipation qualitatively change the nature of
the traversal times near the resonances. Coupling to the slab plasmon polaritons will be shown to give rise to large
delay times. We find that for pulses with large enough frequency bandwidth, the traversal times are positive, 
and subluminal.

\section{Traversal time across a dispersive slab}

We calculate the propagation times for electromagnetic pulses across a dispersive slab. 
The relative dielectric permittivity and the magnetic permeability of the slab are considered to be plasma like and
Lorentz dispersive in nature, and given by,
\begin{equation}
\varepsilon(\omega)=1-
\frac{\omega_p^2}{\omega(\omega +i \gamma_p)},
\end{equation}

\begin{equation}
\mu(\omega) = \frac{\omega^2-\omega_b^2+ i\omega \gamma_m}{\omega^2- \omega_{0}^2 + i\omega \gamma_m}.
\end{equation}
Here $\omega_{p,b,o}$=2$\pi f_{p,b,o}$ with $f_p$, $f_b$, and $f_0$ taken to be 12 GHz, 6 GHz and 4GHz respectively.
The slab behaves as a positive refractive index medium (PRM) or right handed medium (RHM) ($\varepsilon>0$, $\mu>0$) when $\omega > \omega_p$, 
a negative refractive index 
medium (NRM or LHM) ($\varepsilon<0$, $\mu<0$) within $\omega_0 < \omega < \omega_b$, and as a barrier ($\varepsilon<0$, $\mu>0$) elsewhere.
Here we note that the expressions of $\varepsilon$ and $\mu$ are similar to those
given by \cite{duttagupta} with the main difference arising due to the introduction of finite amounts of 
damping in the medium.

\begin{figure}[tb]
\begin{center}
\includegraphics[angle=-0,width=0.90\columnwidth]{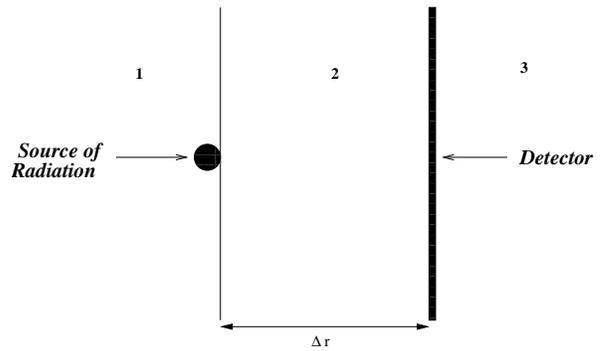}
\end{center}
\caption {A dispersive slab having thickness $\Delta r$ (Region 2) surrounded by vacuum (Regions 1, and 3).
The black circle in Region 1 just outside the slab represents the source of radiation. The black screen in Region 3
just outside the slab boundary represents the detector.} 
\label{plot1}
\end{figure}

For convenience, we take the 
source of radiation to be placed in vacuum just outside one boundary of the slab and the 
detector just outside the other one (Fig. 1). We have taken same medium (vacuum) on either sides of the slab (i.e. regions 1, and 3). 
So here $\varepsilon_{1}=\mu_{1}=\varepsilon_{3}=\mu_{3}=1$. $\varepsilon_{2}$, and $\mu_{2}$ have same forms
respectively given by Eqs. (4), and (5). We take our initial pulse form to be Gaussian in time which is represented by,
\begin{equation}
\mathbf{E}(\mathbf{r} _i=0,t) = \hat{x} \mathbf{E} _0\exp[-\frac{t^2}{\tau^2}] \exp(-i\bar{\omega}t).
\end{equation}
Hence the Fourier transform of the pulse is given by,

\begin{equation}
\mathbf{E} (\mathbf{r} _i,\omega) = \hat{x} \frac{\mathbf{E} _0}{2\sqrt{2}}\tau
e^{-\frac{(\omega - \bar {\omega})^2}{4}\tau^2},
\end{equation}
where $\bar {\omega}$ is the carrier frequency and $\tau$ is the pulse duration.
The magnetic field is simply obtained using the Maxwell's equations,
\begin{equation}
\mathbf{H} (\mathbf{r} _i,\omega) = \hat{y} \frac{\mathbf{E} _0}{2\sqrt{2}}\frac{k_{z1}}{\omega\mu_1\mu _0}\tau
e^{-\frac{(\omega - \bar {\omega})^2}{4}\tau^2},
\end{equation}
where $k_{z1}$ represents the wave vector in the first medium.
We consider P-polarized light for normal incidence. It should be noted that for normal incidence,
the results are independent of the state of polarization of radiation whereas they are dependent  
in case of oblique incidence. Hence for the latter case, we will deal with both the S- and the P- polarizations separately. 
For the P-polarization, the magnetic field at the detector is related to that at the source
via the transmission coefficient across the slab. The final magnetic and the electric fields at the detector are given by
\begin{equation}
\mathbf{H} (\mathbf{r} _f,\omega) = \hat{y} \frac{\mathbf{E} _0}{2\sqrt{2}}\frac{k_{z1}}{\omega\mu_1\mu _0}\tau
e^{-\frac{(\omega - \bar {\omega})^2}{4}\tau^2}\mathbf{T(\omega)},
\end{equation}
and 
\begin{equation}
\mathbf{E} (\mathbf{r} _f,\omega) = \hat{x} \frac{\mathbf{E} _0}{2\sqrt{2}}\frac{k_{z1}k_{z3}}{\omega^2\mu_1\varepsilon_3}
c^2\tau e^{-\frac{(\omega - \bar {\omega})^2}{4}\tau^2}\mathbf{T(\omega)},
\end{equation}
where $k_{z3}$ represents the wave vector in the third medium. 

Here $\mathbf{T(\omega)}$ represents the transmission coefficient across the slab which is given by,
\begin{equation}
\mathbf{T(\omega)} =4\frac{tt' e^{ik_{z2}\Delta r}}{1-r'^2 e^{2ik_{z2}\Delta r}},
\end{equation}
where $\Delta r$ represents the slab thickness and $k_{z2}$ represents the wave vector inside the dispersive slab.
$t$, $t'$, and $r'$ respectively represent the Fresnel coefficients 
of transmission, and reflection by the slab interfaces and are given by,
\begin{eqnarray*}
t=\frac{2\frac{k_{z1}}{\varepsilon_1}}{\frac{k_{z1}}{\varepsilon_1}+\frac{k_{z2}}{\varepsilon_2}},
t'=\frac{2\frac{k_{z2}}{\varepsilon_2}}{\frac{k_{z2}}{\varepsilon_2}+\frac{k_{z3}}{\varepsilon_3}},
r'=\frac{\frac{k_{z2}}{\varepsilon_2}-\frac{k_{z3}}{\varepsilon_3}}
{\frac{k_{z2}}{\varepsilon_2}+\frac{k_{z3}}{\varepsilon_3}}.
\end{eqnarray*}
Here the unprimed, and primed coefficients stand respectively for the coefficients across the first and the second 
boundaries. For S-polarization, in the expressions of the Fresnel coefficients, the $\varepsilon$'s are simply replaced 
by $\mu$'s. Also the Fresnel coefficients relate the electric fields across the interface rather than the magnetic fields.
Suffixes 1, 2, and 3 respectively represent the parameters at the source, slab, and the detector sides as described earlier. 
For convenience we later substitute equal material parameters on both (source and detector) sides of the slab.
We calculate the delay times for different bandwidths of the pulses. For this we use the same notation for the 
broad and the narrowband pulses as discussed in our earlier paper \cite{lipsa}. The pulse has a broadband when  
$\bar{\omega} \tau$=10 or less and it has a narrowband when $\bar{\omega} \tau$= 100 or more.    

The Wigner delay time was calculated using Eq. (11), and is given by,
\begin{equation}
\tau_{\omega}=\frac{\partial \phi}{\partial \omega}=\frac{\frac{\partial p}{\partial \omega} \tan(k_{z2}\Delta r)+ p \sec^2(k_{z2}\Delta r)\frac{\partial k_{z2}}{\partial 
\omega} \Delta r}{1 + p^2 \tan^2(k_{z2} \Delta r)}, 
\end{equation}
where 
\begin{equation}
p = \frac{\frac{k_{z1}\varepsilon_2}{k_{z2}\varepsilon_1}+\frac{k_{z2}\varepsilon_1}{k_{z1}\varepsilon_2}}{2}
\end{equation}
for P-polarization and
\begin{equation}
p = \frac{\frac{k_{z1}\mu_2}{k_{z2}\mu_1}+\frac{k_{z2}\mu_1}{k_{z1}\mu_2}}{2}
\end{equation}
for S-polarization. Throughout our calculations (both for normal and oblique incidence), we have checked that the Wigner 
delay time yields the same result as the traversal time for narrowband pulses with the average energy flow method.  

\subsection{Traversal times for normal incidence}

\begin{figure}[tb]
\begin{center}
\includegraphics[angle=-0,width=0.90\columnwidth]{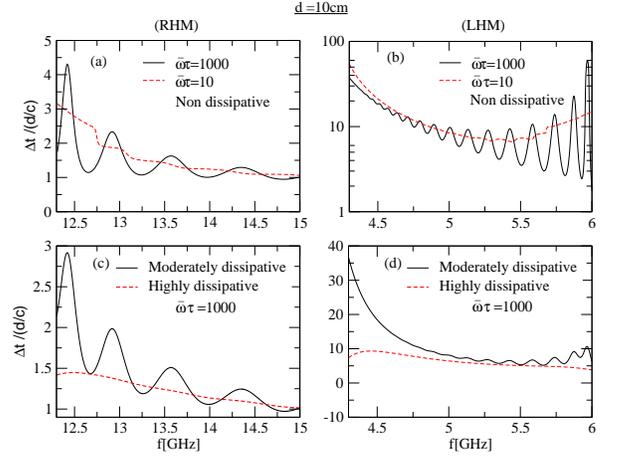}
\end{center}
\caption {(Color online) Scaled total delay time of a pulse with different bandwidths ($\bar{\omega} \tau$) plotted as a 
function of frequency (f) across a dispersive slab with large thickness ($\Delta r =10cm$).
(a) Solid, and dashed lines show delay times respectively for a narrowband, and a broadband pulse across a 
nondissipative slab of RHM.
(b) Same as (a), but across a slab of LHM.
(c) Delay time for a narrowband pulse across a moderately dissipative ($\gamma_p=0.01\omega_p$, $\gamma_m=0.01\omega_b$) 
slab (solid line), and a highly dissipative ($\gamma_p=0.1\omega_p$, $\gamma_m=0.1\omega_b$)
slab (dashed line) of RHM. (d) Same as (c), but across a slab of LHM.}
\label{plot2}
\end{figure}

\begin{figure}[tb]
\begin{center}
\includegraphics[angle=-0,width=0.90\columnwidth]{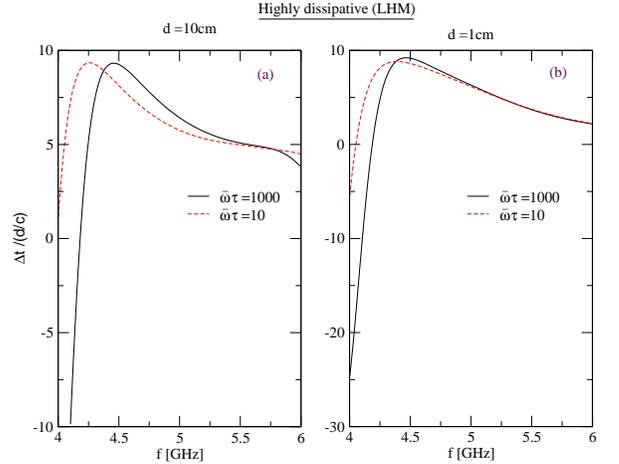}
\end{center}
\caption {(Color online) (a) Scaled total delay time as a function of the carrier frequency (f) for a 
narrowband pulse (solid line), and a broadband pulse (dashed line) across a highly 
dissipative ($\gamma_p=0.1\omega_p$, $\gamma_m=0.1\omega_b$) slab of LHM, and 
large thickness ($\Delta r=10cm$). (b) same as (a), but with a small slab thickness ($\Delta r=1cm$).}
\label{plot3}
\end{figure}

In this case, the parallel component of the wave-vector is zero and the pulse 
is normally incident on the slab. So there is no coupling with the slab plasmon polaritons. Since $k_x=0$, 
the Maxwell's equations can be combined to give,
\begin{equation}
k_z^2=\frac{\omega^2\varepsilon\mu}{c^2}.
\end{equation}
This is independent of whether the slab is of a RHM or a LHM. 
In Fig. 2, we plot the delay times scaled with the free space propagation ($\frac{\Delta r}{c}$) versus the carrier frequency
for both broadband ($\bar{\omega} \tau$= 10) and narrowband ($\bar{\omega} \tau$= 1000) pulses.
 For narrowband pulses, we refer to \cite{duttagupta} where
the Wigner delay times were calculated for a nondissipative slab and it was shown that 
resonant features appear in  the delay time behaviors due to presence of the poles of the transmission 
coefficient (Fabry-P\'{e}rot resonances).
We have taken ($\Delta r =10$cm) as large thickness and ($\Delta r =1$cm) as small thickness 
of the slab relative to the wavelength (2.5cm) of the pulse at the electrical plasma frequency ($f_p$) . 

First we compute the results for the traversal times of broadband 
and narrowband pulses through a nondissipative slab which is achieved by substituting 
$\gamma_p = \gamma_m = 0$ in the expression of 
$\varepsilon$ and $\mu$. Figs. 2(a) and 2(b) respectively show the traversal times for pulse propagation inside slabs with positive and negative refractive indices. In both the figures, it can be observed that, 
the features due to the slab resonances get smoothened with an increase in the pulse bandwidth. 
So it is expected that for extremely broadband light, 
these features might completely disappear. Here we note that, the results for the narrowband pulses in Figs. 2(a), and 2(b),
are exactly same as those for the Wigner delay times given in \cite{duttagupta}. 
Next we study the traversal times for narrowband pulses
propagating through dissipative slabs of both RHM and LHM (Figs. 2(c), and 2(d)). 
To include moderate levels of dissipation in the medium, 
we use $\gamma_p=0.01\omega_p$ and $\gamma_m=0.01\omega_b$ and for high levels of dissipation, we use 
$\gamma_p=0.1\omega_p$ and $\gamma_m=0.1\omega_b$ respectively in Eqs. (4) and (5). 
We see that when a small amount of dissipation is introduced in the medium, the time taken for transmission through the 
slab is less than that taken for the nondissipative case. With increased dissipation in the 
slab, one can also clearly observe that the slab resonant features just disappear.
We have also studied the delay times for a highly dissipative slab of 
LHM for both narrow and broadband pulses (Figs. 3(a), and 3(b)).
Fig. 3(a) shows the results for a slab with large thickness (10cm) and 3(b)shows the corresponding results for a slab with a small 
thickness (1cm). It can be clearly seen that the delay time is very small near the magnetic resonance frequency. Then it 
rapidly increases for large frequencies, and after passing through a peak, it gradually decreases. 
For narrowband pulses, the total delay time near the 
resonance frequency ($\omega_{0}$) even becomes largely negative. Even for broadband pulses, with small thickness
of the slab, this negativity in the delay time appears near $\omega_{0}$ although to smaller and smaller
extent with increasing thickness of the slab. 
The anomalous dispersion of the refractive index of a medium with high amount of dissipation leads to 
small/negative delay times for broadband/narrowband pulses near the resonance frequency. 
It can be seen that, for narrowband pulses, and large thickness of the slab, 
the delay time bends down near $\omega_{b}$ where the value of $\mu$ becomes 0.
For the broadband pulse, the traversal time remains unaffected at $\omega_{b}$.   

\begin{figure}[tb]
\begin{center}
\includegraphics[angle=-0,width=0.90\columnwidth]{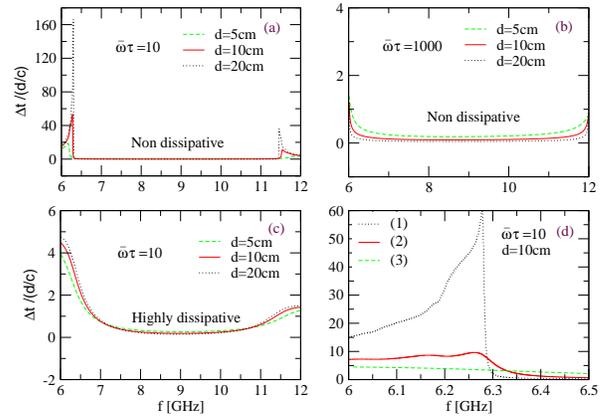}
\end{center}
\caption {(Color online) (a) The scaled total delay time of a broadband pulse as a function of the carrier frequency (f)
across a nondissipative slab which behaves as a plasma medium with different thicknesses shown by 
dashed line ($\Delta r=5cm$), solid line ($\Delta r=10cm$), and dotted line ($\Delta r=20cm$).
(b) same as (a), but the pulse has a narrow bandwidth. (c) same as (a), but the slab is 
highly dissipative ($\gamma_p=0.1\omega_p$, $\gamma_m=0.1\omega_b$).
(d) Fabry P\'{e}rot resonance structures near resonance frequency in the delay time graph for a broadband pulse across 
a slab with large thickness ($\Delta r$=10cm). Dotted, solid, and the dashed lines indicated by numbers 1, 2, and 3
respectively show the features across slabs of nondissipative ($\gamma_p=0$, $\gamma_m=0$),
moderately dissipative ($\gamma_p=0.01\omega_p$, $\gamma_m=0.01\omega_b$), 
and highly dissipative ($\gamma_p=0.1\omega_p$, $\gamma_m=0.1\omega_b$) media.}
\label{plot4}
\end{figure}

\begin{figure*}[tb]
\begin{center}
\includegraphics[angle=-0,width=1.6\columnwidth]{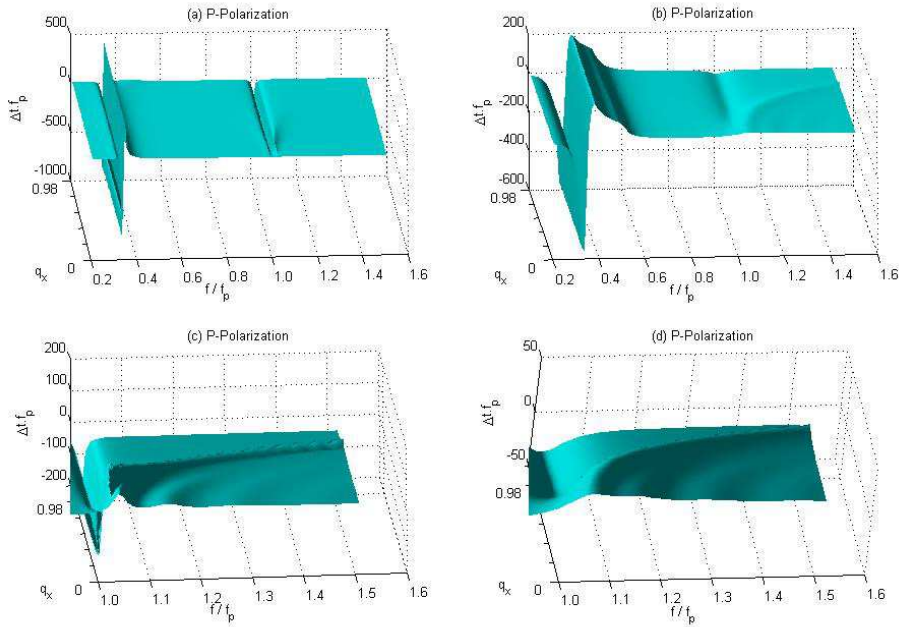}
\end{center}
\caption {(Color online) Total delay time of an extremely narrowband pulse ($\bar{\omega} \tau = 5000$) 
plotted versus scaled carrier frequency ($f/f_p$), and scaled parallel wave vector ($q_x$) for P-polarization across a 
moderately dissipative ($\gamma_p=0.01\omega_p$, $\gamma_m=0.01\omega_b$) slab
with small thickness ($\Delta r=\frac{\lambda_p}{5}$). (b) same as (a), but across a slab of highly dissipative 
($\gamma_p=0.1\omega_p$, $\gamma_m=0.1\omega_b$) material and large thickness ($\Delta r= 2{\lambda_p}$).
(c) Fabry P\'{e}rot resonance structures in the delay time plotted versus scaled carrier frequency, and parallel
wave vector across a moderately dissipative ($\gamma_p=0.01\omega_p$, $\gamma_m=0.01\omega_b$)
slab with large thickness ($\Delta r= 2{\lambda_p}$).
(d) same as (c), but the slab is highly dissipative ($\gamma_p=0.1\omega_p$, $\gamma_m=0.1\omega_b$). }
\label{plot5}
\end{figure*}

Next we investigate the delay times within the frequency range ($\omega_b < \omega < \omega_p$) where the medium behaves as
a plasma and as a consequence, most of the wave components are evanescent. 
We study the traversal times for both broadband and narrowband 
pulses. We find that for different slab thicknesses ($\Delta r=$5cm, $\Delta r=$10cm, $\Delta r=$20cm) involved, and
regardless of the amount of dissipation in the slab material, the traversal time remains the same for almost all frequencies 
except at the edges of the stop bands (Figs. 4(a), 4(b), and 4(c)). 
Such an independent behavior of the traversal time with distance as well as dissipation  
is the famous Hartman effect \cite{hartman}. We also see Fabry-P\'{e}rot resonance-like structures with large peaks 
appearing near the edges of the stop bands frequencies for broadband pulses with large thickness and nondissipative
slab material (Fig. 4(d)).
Such resonant features do not appear if the slab thickness is reduced, or interestingly, if the bandwidth of the pulse is
small either. We can understand this surprising resonant effect for broadband pulses by noting that the plasma acts as a 
spectral filter. Consider a pulse with a large bandwidth and the carrier frequency within the stopband but near the lower 
edge. In transmission, most of the spectral components within stopband are almost completely attenuated while the spectral 
components in the passband at lower frequencies will get transmitted. If the bandwidth of these propagating spectral 
sub-components is small, then the behavior of the transmission as a function of frequency will be modulated by 
Fabry-P\'{e}rot resonances. Note that the impedance mismatch between the slab and vacuum will be much larger at
$\omega_b$, thus emphasizing the resonant effects. Such resonant effects disappear for small slab thickness because 
there the bulk of the spectral components within the stopband are not attenuated enough for the Fabry-P\'{e}rot resonances 
to dominate. Increasing dissipation also obviously destroys the contribution of these resonances to the phenomena. It should
be noted that the transmitted pulses will be significantly stretched or deformed as their bandwidth will be sbstantially 
smaller. We have not studied here this aspect via the deviation $(\Delta \tau)^2$ of the delay time.
\subsection{Traversal times for oblique incidence}

Next, we study the traversal time of a pulse when it is obliquely incident on a slab, i.e., the 
parallel wave vector ($k_x$) is nonzero. For this case, the Maxwell equations give,
\begin{equation}
k_z=\sqrt{\frac{\omega^2\varepsilon\mu}{c^2}-k_x^2}=\frac {\omega}{c}\sqrt{\varepsilon \mu -q_x^2},
\end{equation}
where $k_x=q_x\frac{\omega}{c}$.

We consider the incident pulses with either S or  P polarizations. For convenience, we have scaled all the 
frequencies with respect to the plasma frequency. So here $f_p=1$, $f_0=0.33$, and $f_b=0.5$.
The thicknesses of the slab used are $\Delta r=\frac{\lambda_p}{5}$ (small thickness), for which 
$k_p \Delta r = 0.4\pi$, and $\Delta r= 2{\lambda_p}$ (large thickness), for which $k_p \Delta r = 4\pi$.
The thicknesses are relative to the wavelength ($\lambda_p$) at the plasma frequency. 
This is important to note that, here for narrowband pulses, 
we use $\bar{\omega}\tau=5000$ (extremely narrowband case) and for 
broadband pulses, $\bar{\omega}\tau=1$ (extremely broadband case). 

\subsubsection{Propagating waves}
\begin{figure*}[tb]
\begin{center}
\includegraphics[angle=-0,width=1.6\columnwidth]{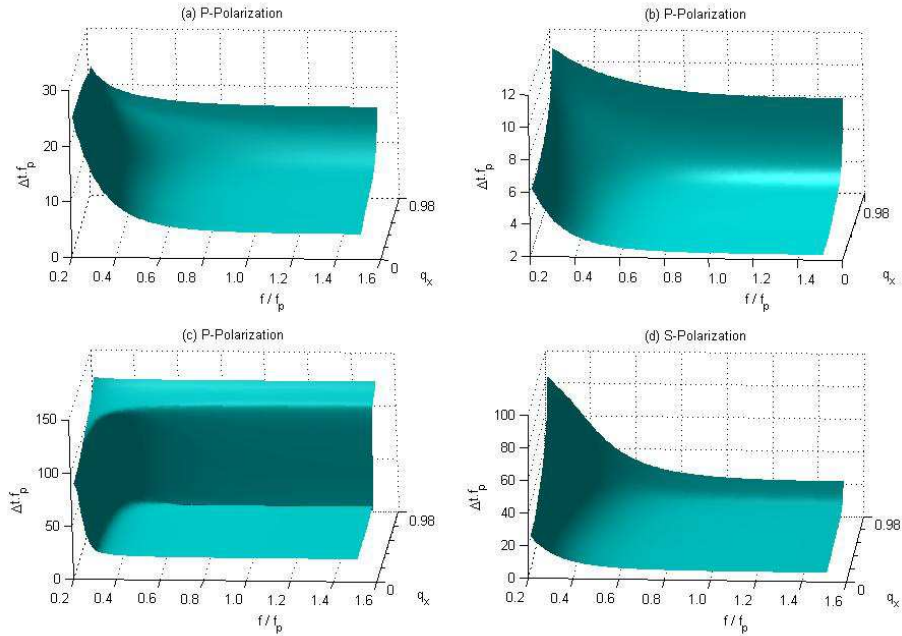}
\end{center}
\caption {(Color online) (a) Total delay time of an extremely broadband pulse ($\bar{\omega} \tau = 1$)
plotted versus scaled carrier frequency ($f/f_p$), and parallel wave vector ($q_x$) across a moderately 
dissipative ($\gamma_p=0.01\omega_p$, $\gamma_m=0.01\omega_b$) slab with small thickness ($\Delta r=\frac{\lambda_p}{5}$),
and P-polarization. (b) same as (a), but the slab is highly dissipative ($\gamma_p=0.1\omega_p$, $\gamma_m=0.1\omega_b$).
(c) same as (b), with large slab thickness ($\Delta r= 2{\lambda_p}$). (d) same as (a), but for S-polarization.}
\label{plot6}
\end{figure*}

Here we discuss the pulses for which the wavevector is real, or in other words, the waves are propagating.
First of all, we present the results for an extremely narrowband pulse ($\bar {\omega} \tau = 5000$)
with moderate amounts of dissipation in the slab material.
The delay time is plotted versus both the frequency, and the parallel wave vector 
(Figs. 5(a), 5(b), 5(c), and 5(d)). From the graphs, it can be clearly 
observed that violent dispersion of the delay time occurs at the magnetic resonant frequency for both
the polarizations. For P- polarization, a small dip in the delay time also occurs at the electric plasma frequency
($f_p$) for small thickness of the slab (Fig. 5(a)), and at both the $f_{p}$, and $f_{b}$ for large 
thickness of the slab. For S- polarization, this additional small dispersion 
in the delay time curve apart from that at $f_0$ occurs only at the magnetic plasma frequency 
where as it becomes smooth at the electrical plasma frequency for all thicknesses of the slab. 
This arises because the Fresnel coefficients for S- polarized waves do not depend on the dielectric 
constants ($\varepsilon$) as strongly as the magnetic permeability ($\mu$).
When there is high amount of dissipation in the material of the slab, 
it is seen that, the scale of the delay time axis as well as the sharpness of the 
dispersion at the resonant frequency decrease simultaneously with the 
broadening of the peaks for both the polarizations. Here we should note that, the traversal times for narrowband pulses 
across slabs with large thickness and high amount of dissipation, give exactly same features with the same scales 
for both the S-, and P- polarizations.  

Then the delay time results are analysed for
narrowband pulses and large thickness of the slab in the 
frequency range which causes positive refractive index. After introducing moderate amount of dissipation in the medium, one can see 
Fabry-P\'{e}rot resonance structures appearing beyond $f_{p}$ (Fig. 5(c)). Such resonant structures (ripples) 
disappear with increase in dissipation in the medium (Fig. 5(d)).

Next we study the traversal times for an extremely broaband pulse ($\bar{\omega}\tau=1$). It can be seen that
the delay times are large at frequencies close to $f_0$ and large wave vectors
(Figs. 6(a), 6(b), 6(c) for P-polarization), and 
(Fig. 6(d) for S-polarization). This happens for all thicknesses of the slab with any amount of dissipation in the 
material. In the delay time graph for S-polarization, a large peak can be seen at frequency close to $f_0$ and large wave 
vector. This occurs due to the large dispersion in $\mu$ at the magnetic resonance and the magnetic plasma frequencies.   
The delay time decreases with increase in dissipation in the medium. It can be observed that the graph 
becomes flat for small thickness of the slab, whereas for large slab thickness,
the delay time increases rapidly at frequencies close to $f_0$ and large wave vectors. For large thickness of the slab with 
high amount of dissipation in the material, both the polarizations give similar features with equal scales (Fig. 6(c) 
for P-polarization). We should note that, for broadband pulses, the total time taken for traversal is 
positive and subluminal for all the parameters studied here.
\begin{figure}[tb]
\begin{center}
\includegraphics[angle=-0,width=0.90\columnwidth]{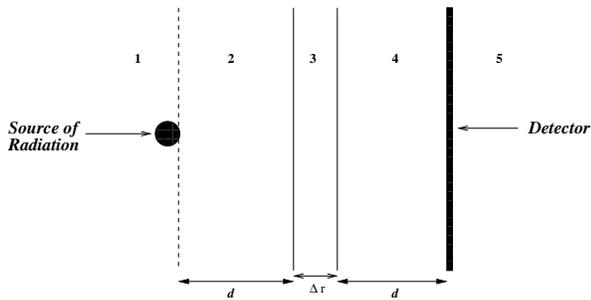}
\end{center}
\caption {A layered structure consisting of five slabs used to study the tunneling of pulses.
Region 3 is the dispersive slab with thickness $\Delta r$
with $\varepsilon_3$, and $\mu_3$ respectively given by Eqs. (4), and (5). Regions 2, and 4 are air slabs
($\varepsilon_2=\mu_2=1$), and with thickness d. Regions 1, and 5 are semi-infinite slabs with
$\varepsilon_1=\varepsilon_5=25$, and $\mu_1=\mu_5=1$. The black circle in Region 1 just outside the boundary of slab 2
represents the source, and the black screen in Region 5 outside slab 4 represents the detector.}
\label{plot7}
\end{figure} 

\subsubsection{Evanescent waves}

Here we consider pulses for which most of the wave vectors are imaginary making the incident wave evanescent. 
This is achieved by making the second term in Eq. (16) under the square root large than the first term. 
We calculated both the Wigner delay time and the Energy delay time for such pulses most of whose components
are evanescent.
While the Wigner delay time can be easily calculated using the phase shifts, calculating the traversal time via the
energy flow for evanescent pulses is a non-trivial problem. This is because the energy flow associated with a single
evanescent wave in vacuum is zero. Thus the poynting vector for pulse with all spectral components having evanescent
wave vectors is zero at the detection point in vacuum. Hence it is not possible to compute the energy traversal
time for evanescent pulses if the slab is embedded in vacuum. One needs to couple the energy in these systems out to measure
the pulses. This is related to the measurement of tunneling quantum particles whereby one needs to raise the particles
with negative energy (evanescent waves) above the barrier before detection. Similarly, we make an arrangement with layered slabs
where the evanescent waves are outcoupled to propagating modes in high-index media.

\begin{figure*}[tb]
\begin{center}
\includegraphics[angle=-0,width=1.6\columnwidth]{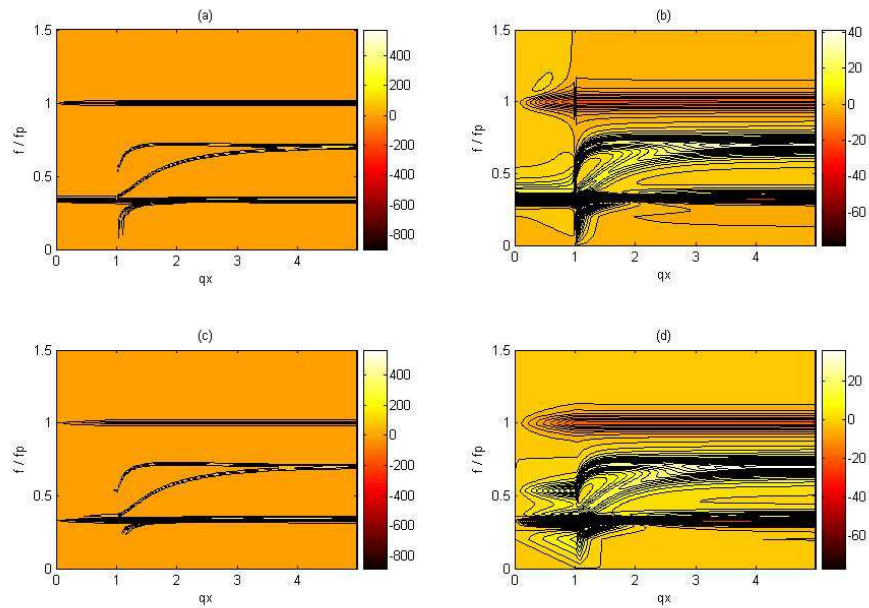}
\end{center}
\caption {(Color online) (a) The Wigner delay times for pulse traversal with P-polarization as a function of frequency
and parallel wave vector across a moderately dissipative ($\gamma_p=0.01\omega_p$, $\gamma_m=0.01\omega_b$)
slab with small thickness ($\Delta r=\frac{\lambda_p}{5}$). (b) Same as (a), but for a highly
dissipative ($\gamma_p=0.1\omega_p$, $\gamma_m=0.1\omega_b$) slab. (c)The traversal times obtained by the energy transport method across the layered
slab structure where the middle dispersive slab is moderately dissipative ($\gamma_p=0.01\omega_p$, $\gamma_m=0.01\omega_b$)
in nature with small thickness ($\Delta r=\frac{\lambda_p}{5}$).
(d) same as (c), but for a highly dissipative ($\gamma_p=0.1\omega_p$, $\gamma_m=0.1\omega_b$) slab.
The dispersion of the slab plasmon polariton modes of the slab
stand out clearly and the resonant conditions for these modes are characterized by large Energy delay times.}
\label{plot8}
\end{figure*}

In this arrangement, we have taken two nondispersive slabs of different parameters kept
symmetrically on either sides of the dispersive slab making a layered slab structure (Fig. 7).
The first and fifth slabs have semi-infinite extent
with large relative dielectric permittivity ($\varepsilon = 25$) and relative magnetic permeability ($\mu=1$).
The second and the 4th slabs consist of vacuum with $\varepsilon=1$, and $\mu=1$ and large slab thickness with corresponding
$k_{p}\Delta r$ equal to (4$\pi$).
The 3rd or the middle dispersive slab has $\varepsilon$, and $\mu$ respectively given by Eqs. (4), and (5) and small slab
thickness with corresponding $k_{p}\Delta r$ equal to (0.4$\pi$).
The source is present in the first medium just outside the boundary of the second slab and the detector is placed
in the fifth medium just outside the boundary of the 4th slab.
The value of $q_x$ in Eq. (16) is chosen in such a manner that the wavevector is real, making the pulse
propagating in first and fifth slabs, and imaginary making the pulse evanescent in second and 4th slabs.

First, we plot the Wigner delay time versus both the frequency and the wave vector in a moderately dissipative slab. 
In a rather uniform landscape of delay times, the resonant conditions for the slab surface plasmon polaritons (SPPs)
stand out in stark contrast where the magnitude of the delay times are comparatively very large. Thus the entire
dispersion of the SPPs can be traced out (Fig. 8). There are two distinct plasmon modes corresponding to the
symmetric and antisymmetric modes whose frequencies tend to
 $\frac{f_p}{\sqrt{2}}$ at large wave vectors. Similarly two modes also appear below the magnetic resonant 
frequency. For highly dissipative slabs also, such plasmon modes are seen for 
evanescent waves, but with large broadening of the dispersion curves (Fig. 8(b)).
The surface plasmon features are lost when the thickness of the slab is larger than $\lambda_p$.
We have seen that the features of the plasmon modes for the S-polarization is different from the P-polarization
as in that case, these modes of magnetic character occur between $f_0$ (magnetic resonant frequency), and $f_{b}$
(magnetic plasma frequency) when $\mu<0$, and superpose at  $\frac{f_{b}}{\sqrt{2}}$.

Then we studied the traversal times for evanescent pulses having narrow bandwidth ($\bar{\omega} \tau = 5000$)
using the energy transport method with our new arrangement of the layered slab structure (Fig. 7).
For this arrangement, we plotted the delay times for narrowband pulses versus both the frequency and the wavevector, 
and analysed the results for moderate and large dissipative slabs with both P-polarization 
(Figs. 8(c), and 8(d)) and S-polarization. 
We see that Figs. 8(c), and 8(d) look almost same as Figs. 8(a), and 8(b). 
Thus, the energy traversal times are also significantly affected at the Surface Plasmon Polariton frequencies.
It is worth noting that the traversal times are large at the resonant conditions.

\subsubsection{Propagation through a slab with $n=-1$}
Finally, we consider a slab having unit negative refractive index ($n=-1$) and surrounded by vacuum ($n=+1$).
Negative refractive index of unit magnitude can be achieved at a single frequency for a nondissipative slab. 
The properties of such a slab with ($n=-1$) are very interesting due to the possibility of designing
a perfect lens \cite{pendry,sar05}
By choosing $f_{p}=1$, $f_{0}=0.33$, $f_{b}=0.5$, we get $n=-1$ at $f=\frac{f_p}{\sqrt{2}}$ ($\varepsilon$=-1, $\mu$=-1). 
With propagation inside the medium, the propagation distance increases by a factor of $\frac{1}{\cos\theta}$.
Here due to perfect impedance matching, no multiple reflections take place.
Using the expression for group velocity ($v_g=\frac{\partial \omega}{\partial k}$), 
the group delay along the direction of propagation is given by
$G_d=\frac{\Delta r}{v_g\cos\theta}=\frac{\Delta r}{c\cos\theta} (n+\omega\frac{\partial n}{\partial \omega})$.
For the particular frequency $f=\frac{f_p}{\sqrt{2}}$, the second term within the bracket in the above expression gives a value equal 
to $\frac{32}{7}$. We plot both the Wigner delay time, and the group delay time versus $q_x$ for propagating pulses for 
the particular frequency mode described above (Fig. 9). From the graph, it can be observed
that the delay time gradually increases with $q_x$ until  $q_x=1$  (where it becomes infinity). 
It can be seen that the graph feature of the group delay time is very similar to the Wigner delay time. 
Hence it is  inferred that the group delay mainly contributes to the total delay occuring
during the propagation of a pulse inside a slab with  $n=-1$. 

\section{Conclusions}
In summary, we have demonstrated the manifold implications of dissipation on the traversal time of a pulse across a 
dispersive slab. Throughout our calculations, we have used average energy flow method \cite{peatross} 
to obtain the delay time of a pulse, and have checked numerically that
the results obtained using the Wigner delay time method, and those using the average energy flow method are exactly same
for very narrow bandwidth pulses. In our results we have shown that high amount of dissipation in slab material, 
along with large pulse bandwidth smoothen out the resonant features.

\begin{figure}[tb]
\begin{center}
\includegraphics[angle=-0,width=0.90\columnwidth]{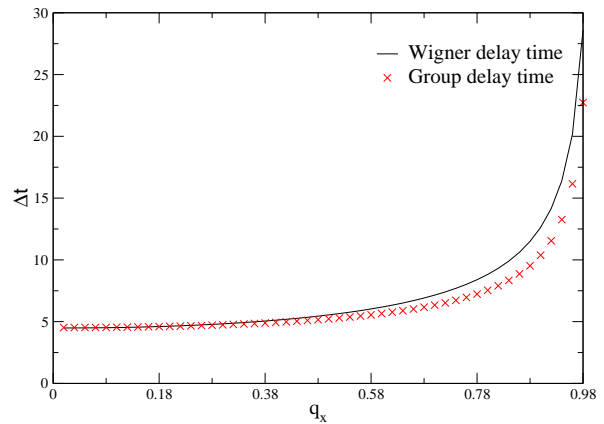}
\end{center}
\caption {(Color online)The Wigner delay time (solid line), and the Group delay time (cross symbol) plotted versus
scaled parallel wave vector($q_x$) across a slab with small thickness ($\Delta r=\frac{\lambda_p}{5}$),
and unit negative refractive index ($n=-1$) surrounded by vacuum.}
\label{plot9}
\end{figure} 
          
We have analysed the reason behind the occurrence of small/negative delay times near the 
magnetic resonant frequency which is a consequence of
anomalous dispersion of the refractive index of the slab medium. We have shown that large slab thickness, along with
high material dissipation give rise to same features for both S and P polarizations in the case of oblique incidence. 
We have also shown that the group delay mainly contributes to the
total delay across a slab with unit negative refractive index, and surrounded by vacuum. 
We have checked that the total time taken for a broadband pulse is always positive, and subluminal.

\section*{Acknowledgement}
S.A.R acknowledges support from the Department of Science and Technology, India under grant no.SR/S2/CMP-54/2003. 
L.N. acknowledges funding from the University Grants Commission, India.

\end{document}